\documentclass{article}
\usepackage{romp}
\usepackage{cite}
\usepackage{graphicx}
\usepackage{psfrag}
\usepackage{amsmath}
\usepackage{hyperref}

\title{Simple cubic random-site percolation thresholds for complex neighbourhoods}
\author{{\L}ukasz Kurzawski and Krzysztof Malarz\thanks{\tt ~~http://home.agh.edu.pl/malarz/}\\ 
	AGH University of Science and Technology,\\
	Faculty of Physics and Applied Computer Science,\\
	al. Mickiewicza 30, 30-059 Krakow, Poland. \\ e-mail: malarz@agh.edu.pl}

\begin{document}

\maketitle
\begin{abstract}
In this communication with computer simulation we evaluate simple cubic random-site percolation thresholds for neighbourhoods including the nearest neighbours (NN), the {next}-nearest neighbours (2NN) and the {next-next}-nearest neighbours (3NN). Our estimations base on finite-size scaling analysis of the percolation probability vs. site occupation probability plots. The Hoshen--Kopelman algorithm has been applied for cluster labelling. The calculated thresholds are 0.1372(1), 0.1420(1), 0.0976(1), 0.1991(1), 0.1036(1), 0.2455(1) for (NN + 2NN), (NN + 3NN), (NN + 2NN + 3NN), 2NN, (2NN + 3NN), 3NN neighbourhoods, respectively. In contrast to the results obtained for a square lattice the calculated percolation thresholds decrease monotonically with the site coordination number $z$, at least for our inspected neighbourhoods.
\end{abstract}


\section{Introduction}

Percolation \cite{ancient,sykes} is a mathematical description of a geometrical phase transition. This allows for purely theoretical studies of all phenomena occurring near the critical point with computer experiments solely or, sometimes, even analytically \cite{anal}. The mixture of occupied and empty sites of given lattice may exhibit some features of real physical systems. Among typical applications of the percolation theory one may find material science \cite{materials}, immunology \cite{immunology} or forest fires problems \cite{forestfires} and studies of liquids moving in porous media \cite{porous}, etc. \cite{Stauffer-Aharony,book}.

Generally speaking, the percolation theory deals with statistical properties of the clusters of occupied nodes (site percolation) or occupied edges (bond percolation) for a given graph, network or regular lattice. In the site percolation problem, the cluster is defined as a group of the occupied lattice vertexes which {are direct or indirect neighbours}. When each site is occupied with some probability $p$ there is a critical probability of sites occupation $p_c$ above which {a} cluster spanning through the whole system appears for the first time \cite{Stauffer-Aharony,book}. This special probability is called percolation threshold $p_c$ and it separates two phases (in the language of material science a conductor and an isolator). The value of percolation threshold $p_c$ depends on kind of percolation (site/bond), lattice/graph/network topology and assumed sites neighbourhoods. In the simplest case only the nearest neighbours constitute the neighbourhoods (von Neumann's neighbourhood) or the nearest neighbours and {next}-nearest neighbours are considered (Moore's neighbourhood).

In the vicinity of the phase transition the quantity $A$ describing the system follows a scaling relation
\begin{equation}
A\propto L^\alpha\cdot f(x \cdot L^\beta),
\end{equation}
where $L$ describes the linear size of the system, $f(\cdot)$ is a scaling function and $x$ is dimensionless scaling parameter \cite{universality}. For physical systems $x$ usually plays the role of reduced temperature $(T-T_C)/T_C$, where $T_C$ stands for critical temperature. The $\alpha$ and $\beta$ parameters are universal exponents which---in the first approximation---do not depend on system details (kind of order/disorder phenomenon, lattice shape, site or bond percolation, etc.) but only on the system dimensionality \cite{Stauffer-Aharony,universality}. However, for $\alpha$ and $\beta$ calculations the precise value of $T_C$ is required. For the geometrical model of the phase transition percolation threshold $p_c$ plays the role of critical temperature.

In this communication we evaluate with computer simulations \cite{kurzawski} the random-site simple cubic percolation thresholds for neighbourhoods including the nearest-neighbours (NN), the second-nearest neighbours (2NN) and the third-nearest neighbours (3NN). Our estimations base on finite size scaling analysis \cite{Stauffer-Aharony,universality} of the percolation probability vs. site occupation probability plots. The Hoshen--Kopelman algorithm \cite{hka} has been applied for cluster labelling. The calculated thresholds $p_c$ are 0.1372(1), 0.1420(1), 0.0976(1), 0.1991(1), 0.1036(1), 0.2455(1) for (NN + 2NN), (NN + 3NN), (NN + 2NN + 3NN), 2NN, (2NN + 3NN), 3NN neighbourhoods, respectively. In contrast to the results obtained for a square lattice \cite{km-sg,sg-km,mm-km} the calculated percolation thresholds decrease monotonically with the site coordination number $z$, at least for our inspected neighbourhoods.

%% ----------------------------------------------------------------------------
\begin{figure*}[!htbp]
(NN)  \includegraphics[width=0.25\textwidth]{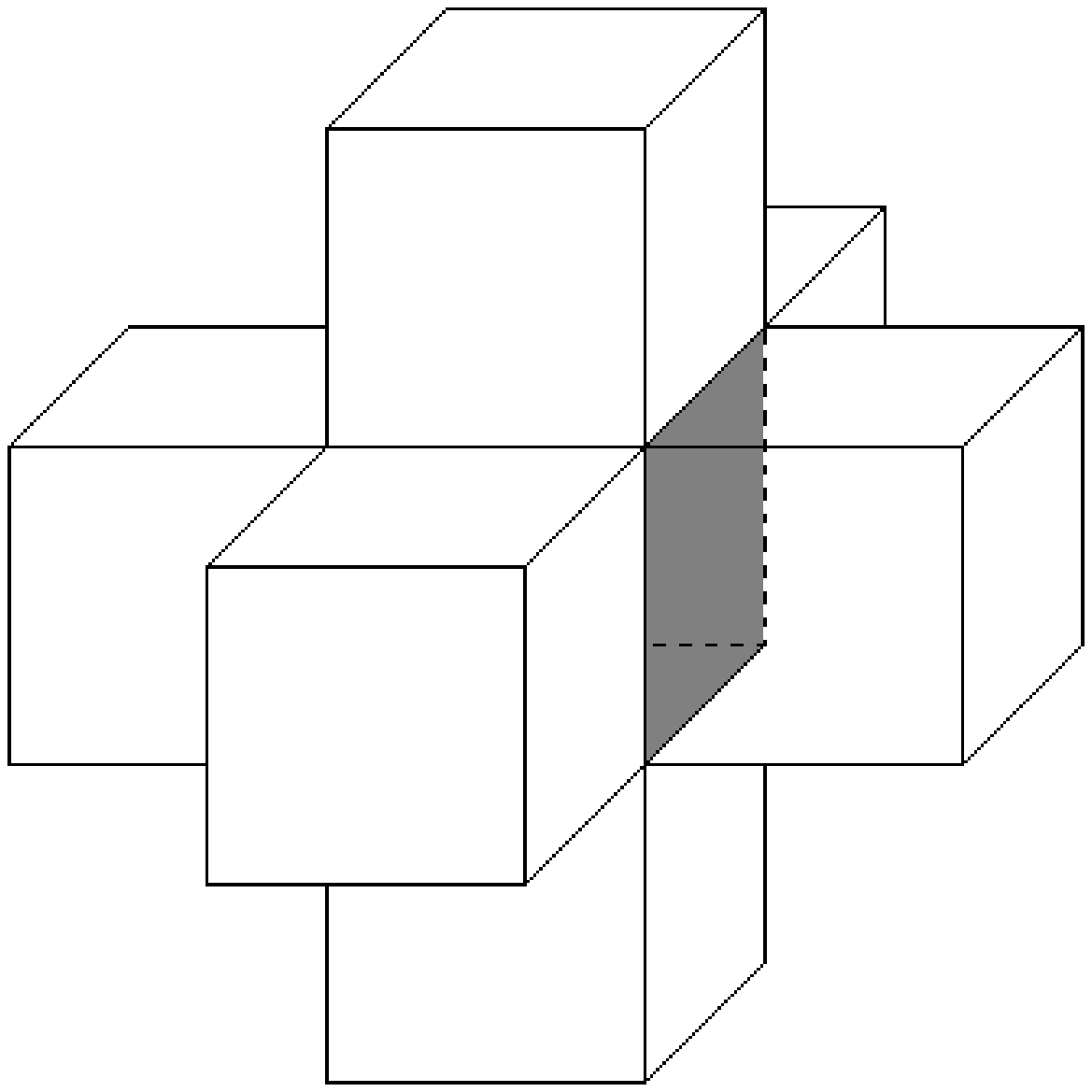}
(2NN) \includegraphics[width=0.25\textwidth]{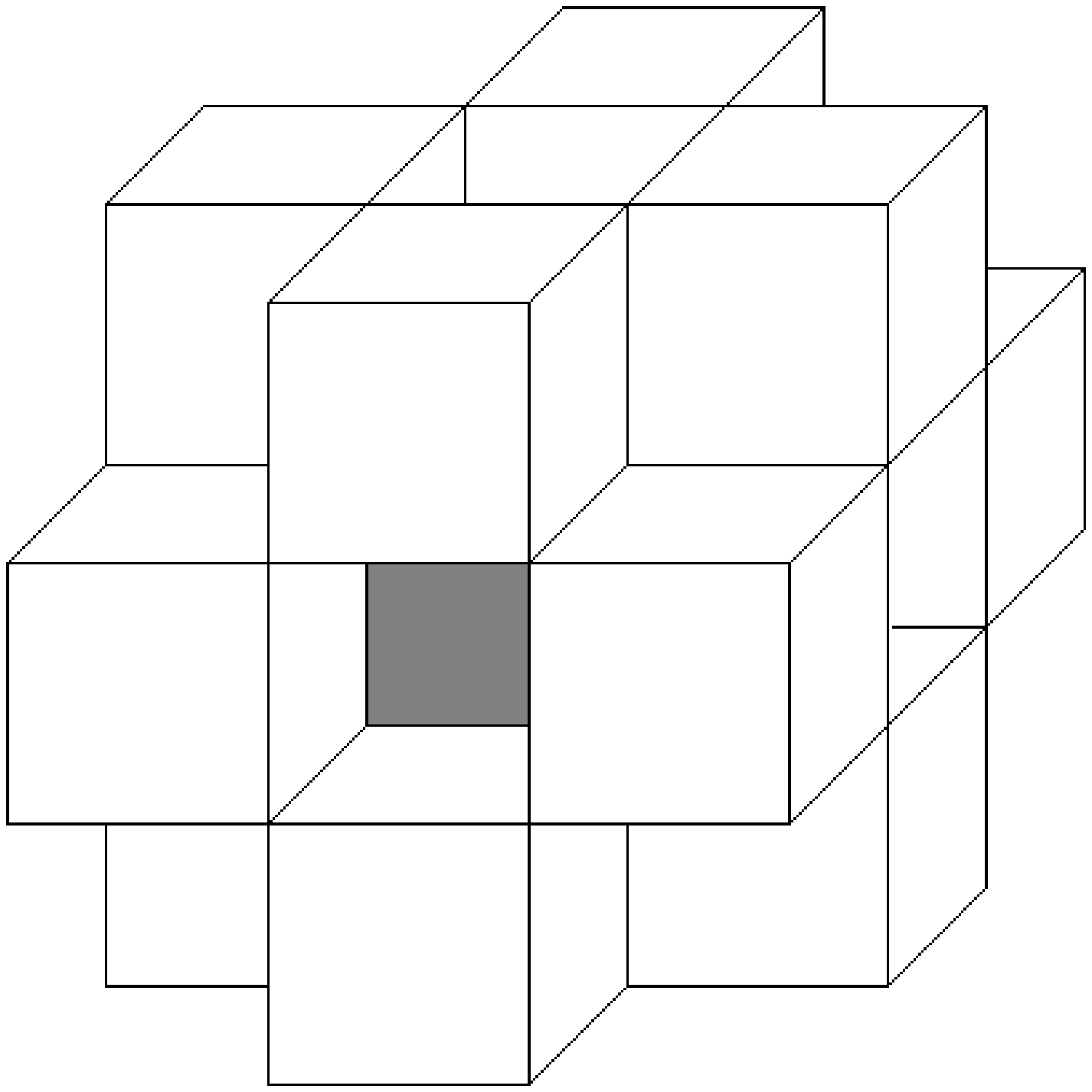}
(3NN) \includegraphics[width=0.25\textwidth]{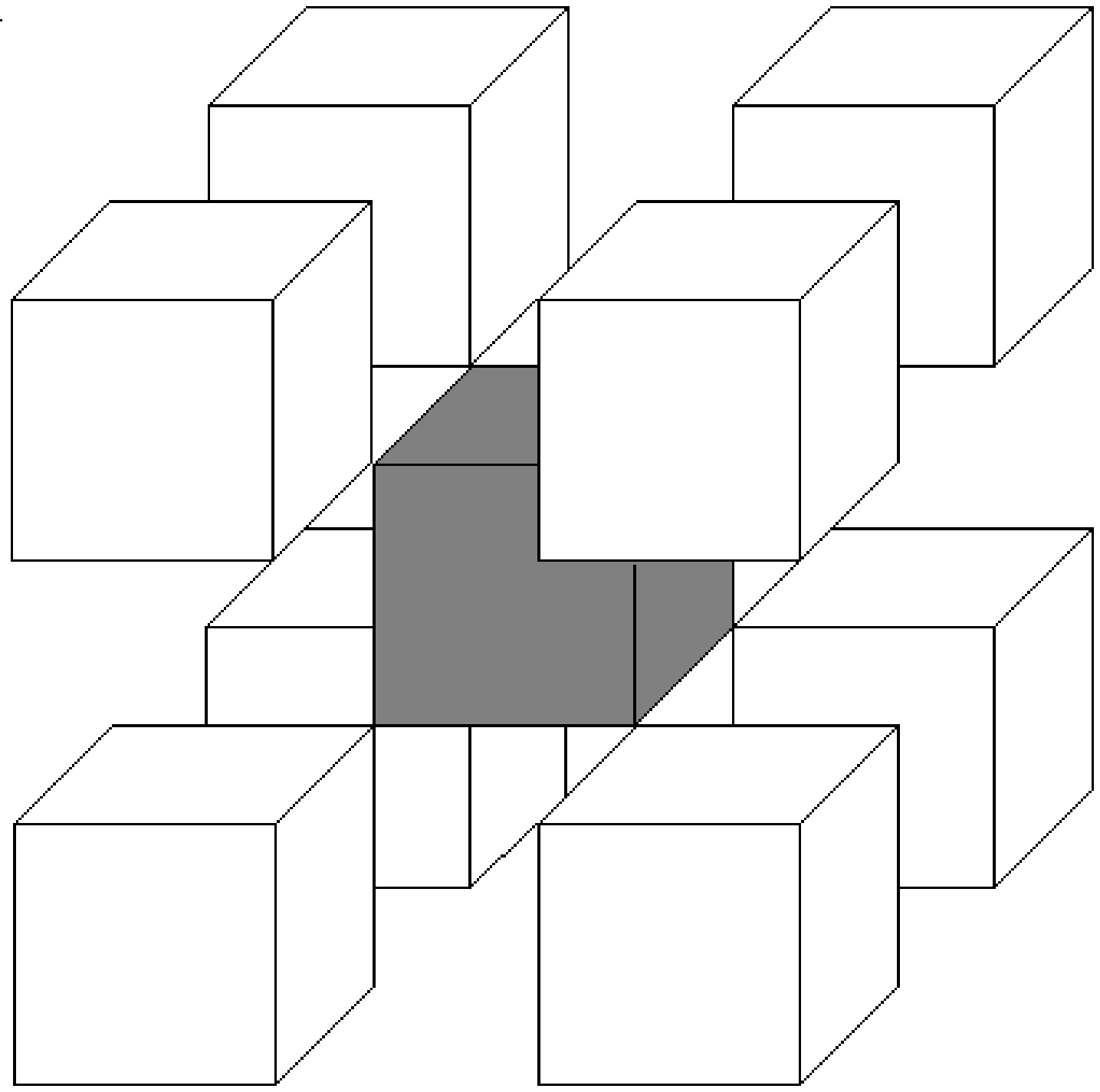}
\caption{Basic neighbourhoods for simple cubic lattice.
The NN neighbourhood is often referred to as von Neumann's neighbourhood while combination of (NN+2NN) is called Moore's neighbourhood.
We propose to name (NN+2NN+3NN) neighbourhood in simple cubic lattice the Rubik's neighbourhood.}
\label{3dnn}
\end{figure*}
%% ----------------------------------------------------------------------------

\section{Calculations}

We use Hoshen--Kopelman algorithm \cite{hka} for occupied sites labelling.
In the Hoshen--Kopelman scheme each site has one label: all sites in a given cluster have the same labels and different clusters have assigned different labels.

\begin{figure*}[htbp]
\psfrag{2NN}{(2NN)}
\psfrag{3NN}{(3NN)}
\psfrag{NN+2NN}{(NN+2NN)}
\psfrag{1NN+3NN}{(NN+3NN)}
\psfrag{2NN+3NN}{(2NN+3NN)}
\psfrag{1NN+2NN+3NN}{(NN+2NN+3NN)}
\psfrag{L, N}{$L, N$}
\psfrag{p}{$p$}
\psfrag{P(p)}{$P(p)$}
\includegraphics[width=0.49\textwidth]{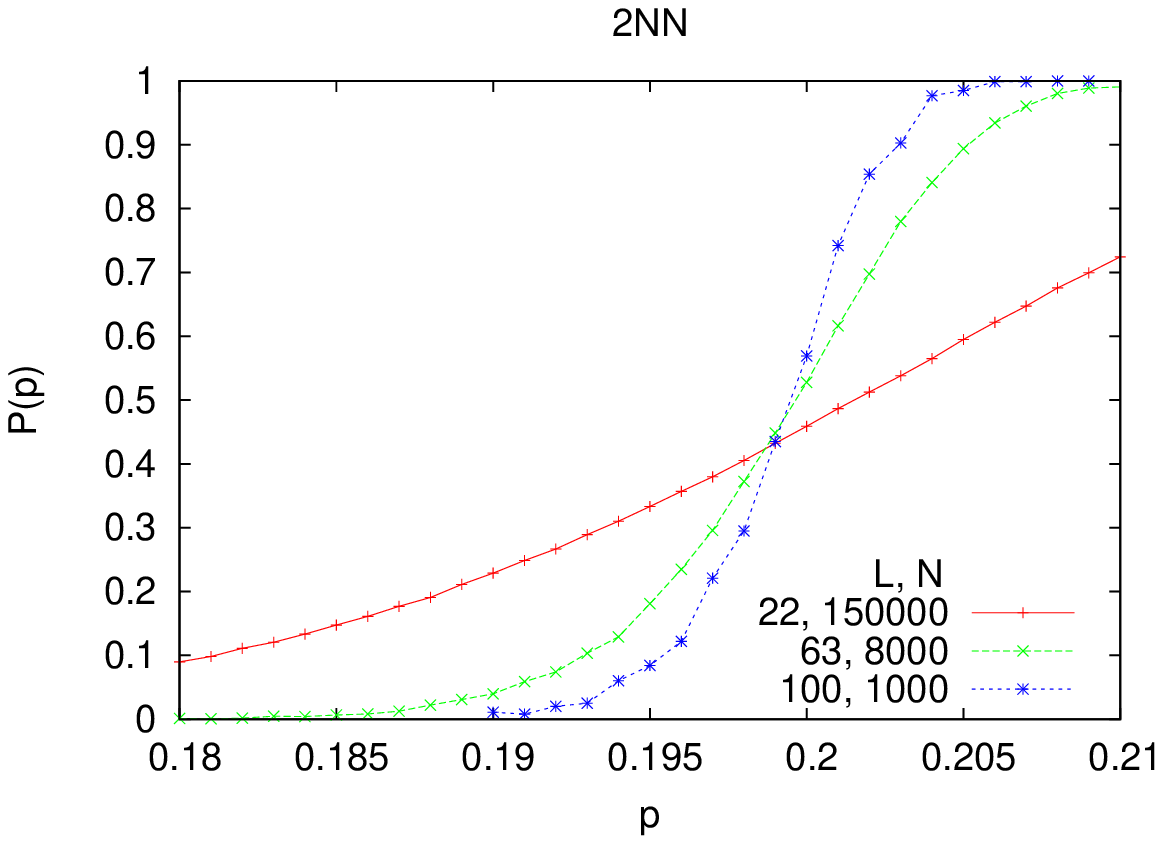}
\includegraphics[width=0.49\textwidth]{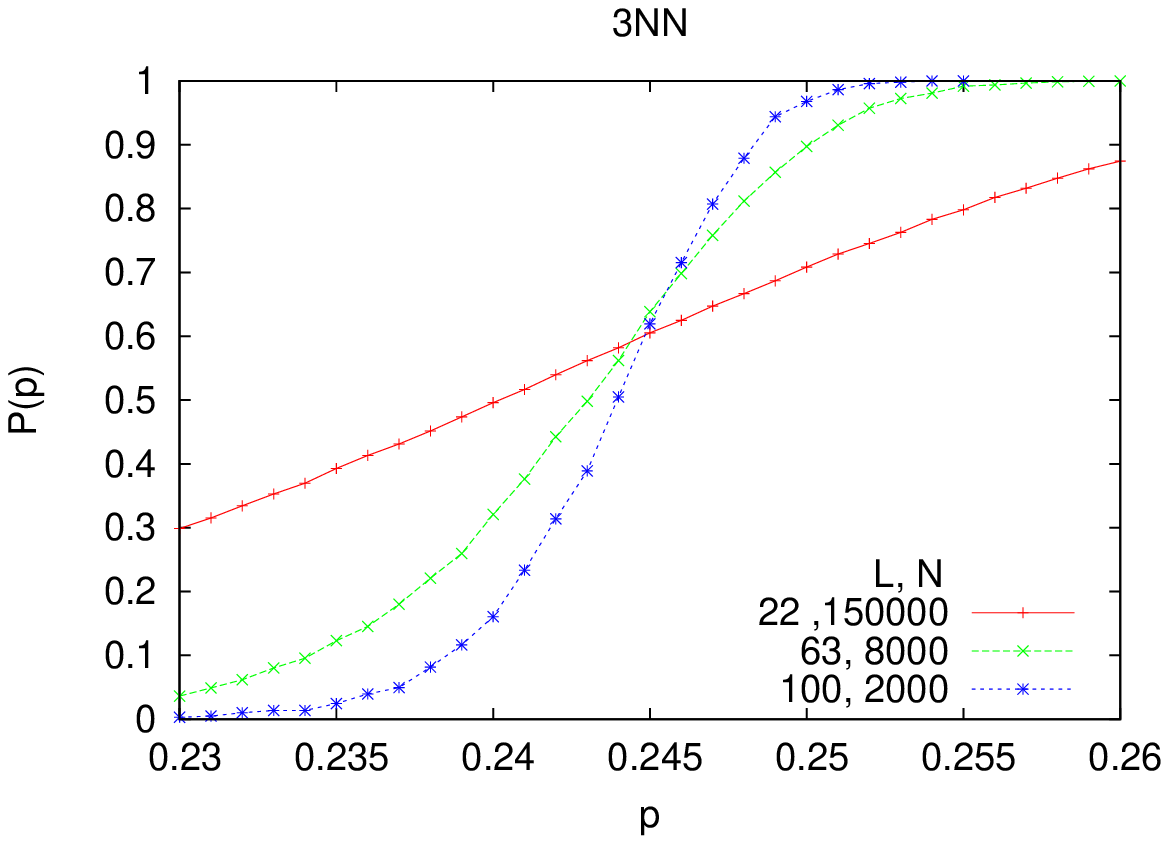}\\
\includegraphics[width=0.49\textwidth]{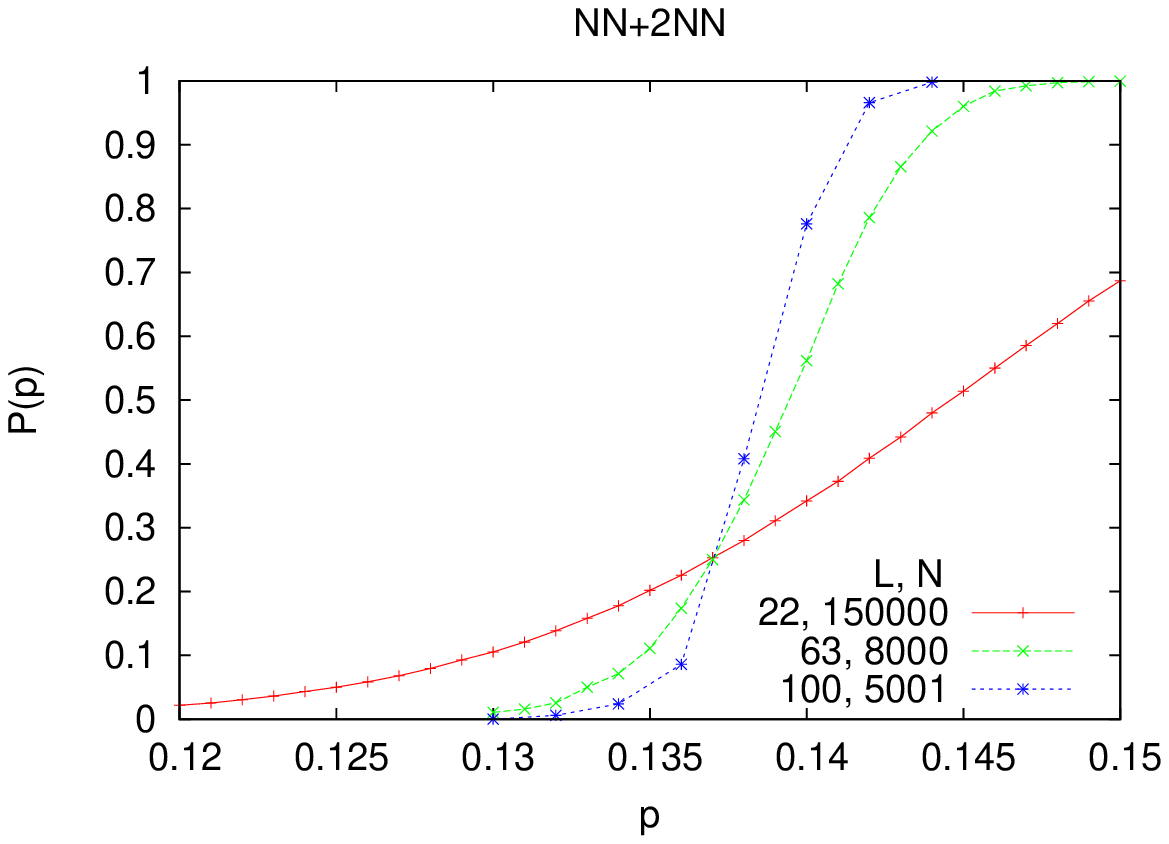}
\includegraphics[width=0.49\textwidth]{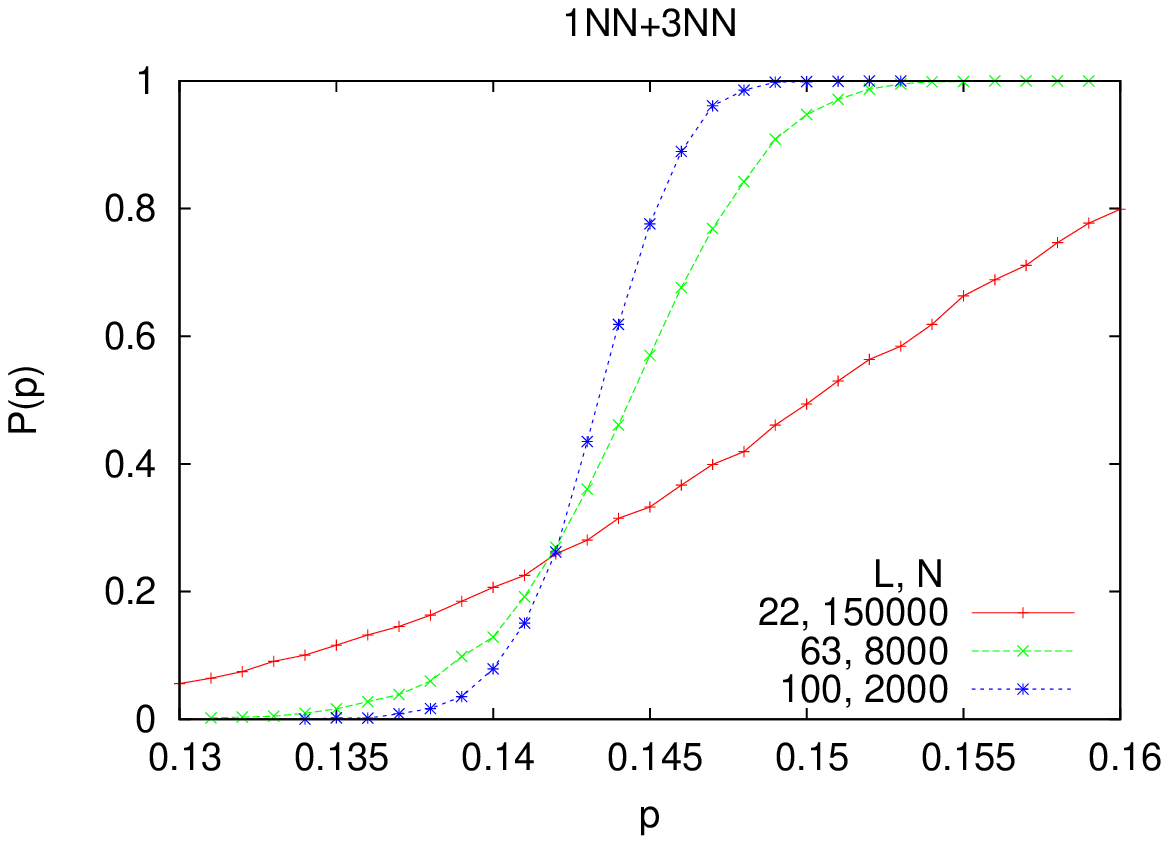}\\
\includegraphics[width=0.49\textwidth]{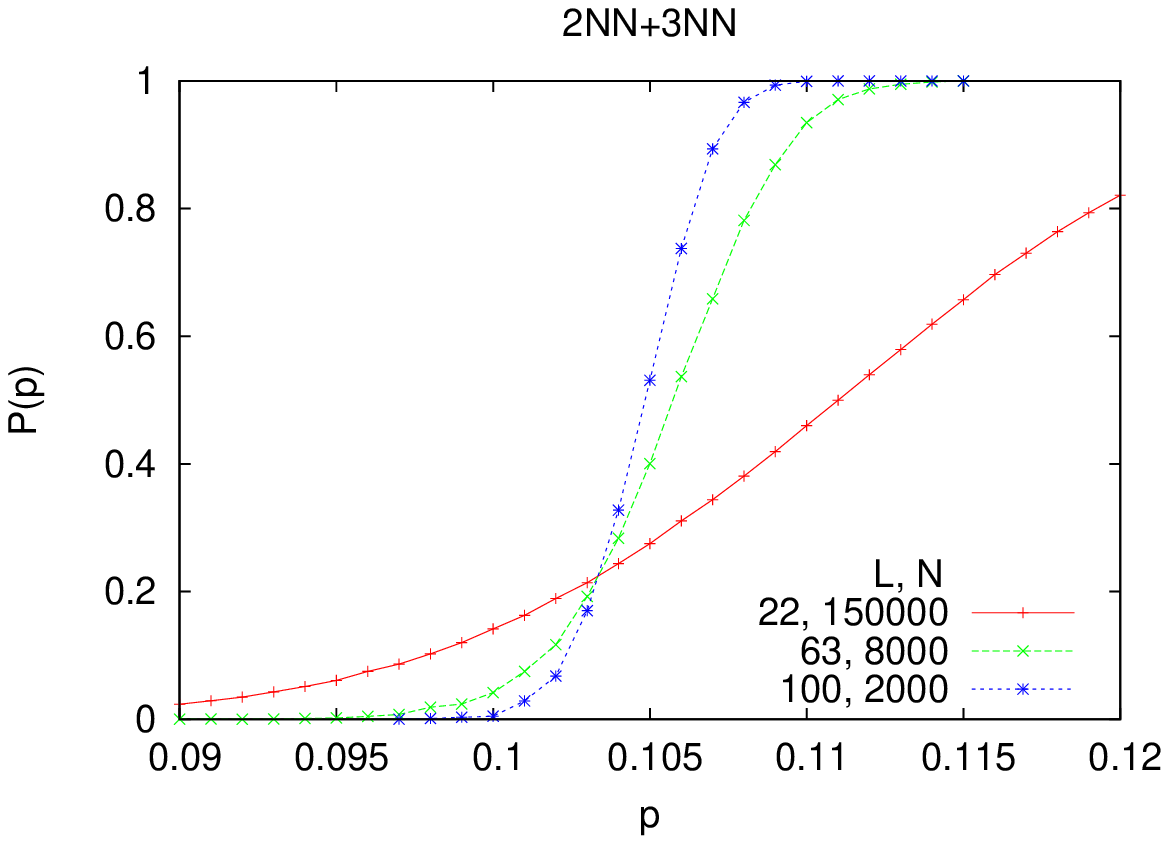}
\includegraphics[width=0.49\textwidth]{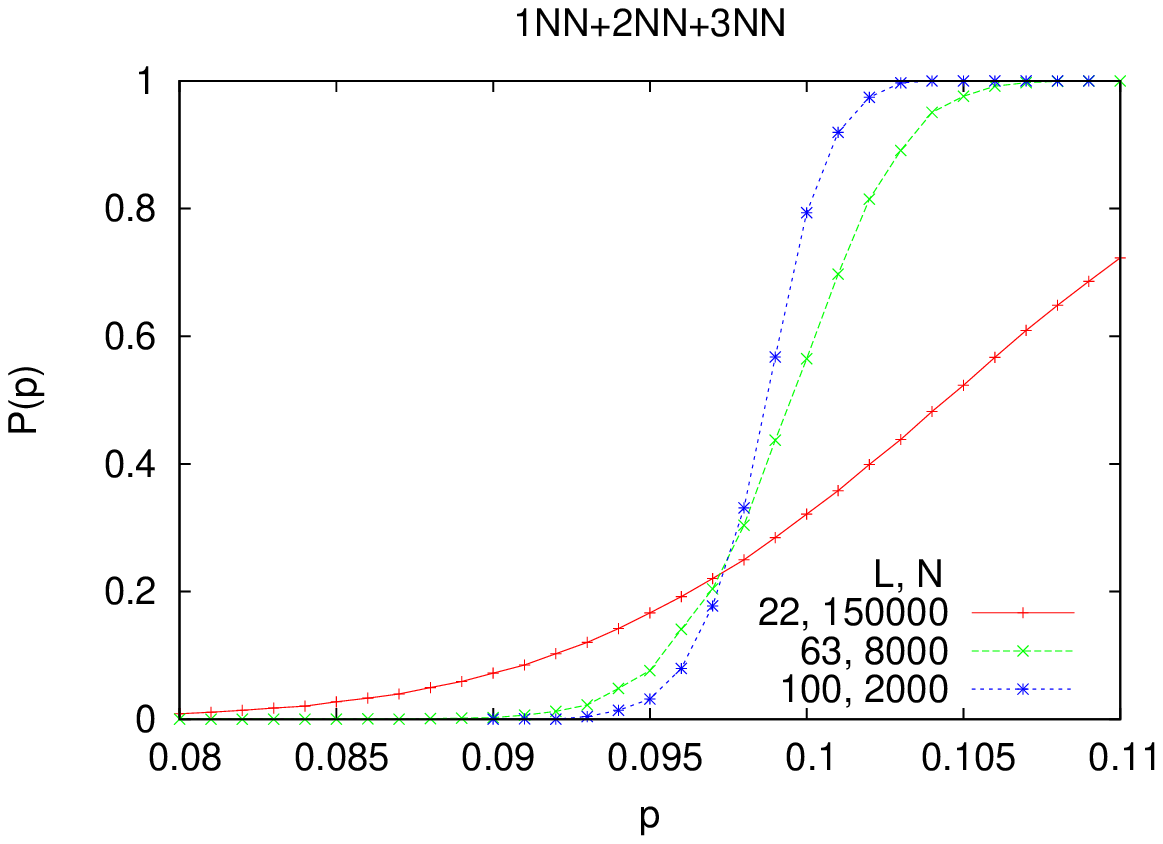}
\caption{Percolation probability $P$ vs. site occupation probability $p$ for various neighbourhoods in three dimensions for three lattice sizes $L$. 
The results presented here have been averaged over $N$ {lattice} realisations.}
\label{P-vs-p}
\end{figure*}

Examples of percolation probability $P$ vs. sites occupation probability $p$ for various neighbourhoods and for various lattice linear sizes $L$ are presented in Fig. \ref{P-vs-p}.
We use finite-size scaling analysis to determine $p_c$ numerically.
As for finite systems the phase transition is never sharp we observe it when for increasing lattice sizes $L$ the $P(p)$ curves become more and more steep and intersect at a common point corresponding also to the case of $L\to\infty$ \cite{privman}.
The common cross-point predicts percolation threshold $p_c$. 
Such strategy was successfully applied for many system description where phase transition may be observed including percolation \cite{newman}, Ising model \cite{IsingTc}, majority-vote models \cite{VM} or opinion dynamics \cite{stepper}.

%% ----------------------------------------------------------------------------
%% ----------------------------------------------------------------------------

Basing on $P(p)$ dependence for various $L$ we look for an interval of the length {$\Delta p=2\cdot 10^{-4}$} where curves for $L=63$ and 100 cross each other.
The results have been averaged over $N=10^5$ lattice realisations.
For example for 2NN case this interval is (0.1990,0.1992) {and the centre of this bracket plays the role of percolation threshold.
Basing on a uncertainty type B evaluation procedure} \cite{typeB} {we estimate the uncertainty $u(p_c)=\Delta p/\sqrt{3}\approx 10^{-4}$.}

%% ############################################################################
\section{Results}
%% ############################################################################

The $P(p)$ dependencies for $L=22$, 63 and 100 are presented in Fig. \ref{P-vs-p}. 
The evaluated percolation thresholds $p_c$ for various neighbourhoods are collected in Tab. \ref{tab-pc}.
To check the accuracy of our estimations we evaluated the percolation threshold for {compact} neighbourhoods as well.
The obtained values $p_c(\text{NN})=0.311$6(1), {$p_c(\text{NN+2NN})=0.1372(1)$ and $p_c(\text{NN+2NN+3NN})=0.0976(1)$} agree {nicely} with the results of extensive numerical simulations \cite{pc-simplecubic} and the earlier estimations \cite{domb-dalton,wiki}.
%% ----------------------------------------------------------------------------
\begin{table}[!ht]
\caption{\label{tab-pc} Simple cubic lattice random-site percolation thresholds $p_c$ for various neighbourhoods constructed with basic neighbourhoods NN, 2NN and 3NN.}
\begin{center}
\begin{tabular}{lrll}
\hline \hline
neighbourhood 	& $z$ 	& $p_c$ & {earlier estimations}\\
\hline
NN              & 6	& 0.3116(1) & {$0.31160\cdots$} \cite{pc-simplecubic} \\ 
2NN             & 12	& 0.1991(1) & \\
3NN             & 8	& 0.2455(1) & \\
NN+2NN    	& 18	& 0.1372(1) & {0.137} \cite{domb-dalton}{, 0.13735(5)} \cite{wiki} \\
NN+3NN   	& 14	& 0.1420(1) & \\
2NN+3NN    	& 20	& 0.1036(1) & \\
NN+2NN+3NN	& 26	& 0.0976(1) & {0.097} \cite{domb-dalton}{, 0.0976445(10)} \cite{wiki} \\
\hline \hline
\end{tabular}
\end{center}
\end{table}
%% ----------------------------------------------------------------------------

%% ----------------------------------------------------------------------------
\begin{figure}[htbp]
\psfrag{p_c}{$p_c$}
\psfrag{z}{$z$}
\begin{center}
\includegraphics[width=0.65\textwidth]{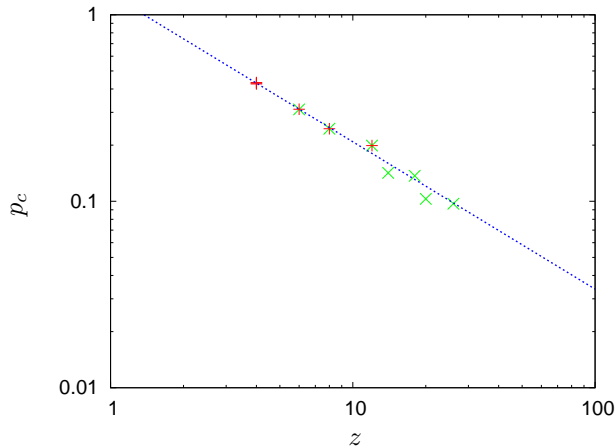}
\end{center}
\caption{\label{pc-vs-z} Percolation thresholds $p_c$ vs. sites coordination number $z$ for {our ($\times$)} inspected three-dimensional neighbourhoods {and other three-dimensional lattices} \cite{ice,diamond,vdSCM,hcp,bcc,bcc+fcc,La2-xSrxCuO4} {estimations ($+$)}. 
The straight lines are the least squares fits of $p_c(z)\propto z^{-\gamma}$ to the experimental data.} 
\end{figure}
%% ----------------------------------------------------------------------------

These $p_c$ values decrease with sites coordination number $z$ as shown in Fig. \ref{pc-vs-z}.
The $p_c(z)$ dependencies may be fitted nicely by {a} straight line in a logarithmic plot, namely: $p_c(z)\propto z^{-\gamma}$ with {$\gamma=0.790(26)$}.
{The percolation thresholds for some three-dimensional lattices have been used for fitting $p_c(z)$ dependence including ice} \cite{ice}{, diamond} \cite{diamond,vdSCM}{, hpc} \cite{hcp}{, bcc} \cite{vdSCM,bcc,bcc+fcc}{, fcc} \cite{vdSCM,bcc+fcc} { and} La$_{2-x}$Sr$_x$CuO$_4$ \cite{La2-xSrxCuO4} {lattices.}
{These additional percolation threshold values are indicated as pluses (+) in Fig.} \ref{pc-vs-z}{.}

%% ############################################################################
\section{Conclusions}
%% ############################################################################
In this communication for the first time we evaluate the random-site percolation thresholds for the simple cubic lattice with {2NN, 3NN, NN+3NN and 2NN+3NN} neighbourhoods for which sites from the first, the second and the third coordination shells were included (see Tab. \ref{tab-pc}). 
The obtained thresholds $p_c$ decrease monotonically with sites coordination number $z$ according to the power law $p_c\propto z^{-\gamma}$, with exponent {$\gamma=0.790(26)$}.

In contrast to the results obtained for a square lattice \cite{mm-km} the calculated percolation thresholds decrease monotonically with the site coordination number $z$, at least for inspected neighbourhoods.

The obtained results may be helpful in studies of the universal formulae \cite{pc-sahimi} for percolation thresholds $p_c$ dependence on sites coordination number $z$.

Finally, we propose to name (NN+2NN+3NN) neighbourhood in simple cubic lattice {\em the Rubik's neighbourhood} as it is identical with the famous Rubik's cube \cite{rubik}---a very popular logical puzzle in early 80's.

%% ============================================================================
\section*{Acknowledgements}
%% \begin{}
{We are grateful to Dietrich Stauffer and an anonymous Referee for paying our attention to Ref.} \cite{domb-dalton} {and Ref.} \cite{wiki}{, respectively.}
Supported by the Polish Ministry of Science and Higher Education and its grants for scientific research.
The numerical calculations were carried out in the Academic Computer Centre CY\-F\-RO\-NET\---AGH (grant{s} No. MEiN/SGI3700/\-AGH/\-024/2006 {and MNiSW/Zeus\_lokalnie/AGH/068/2011}).
%% \end{acknowledgments}
%% ============================================================================

%% ############################################################################

\end{document}